\documentclass{article}

\usepackage{arxiv}

\usepackage[utf8]{inputenc} 
\usepackage[T1]{fontenc}    
\usepackage{hyperref}       
\usepackage{url}            
\usepackage{booktabs}       
\usepackage{amsfonts}       
\usepackage{nicefrac}       
\usepackage{microtype}      
\usepackage{lipsum}		
\usepackage{graphicx}
\usepackage{natbib}
\usepackage{doi}

\title{Linear mixed models for complex survey data: implementing and evaluating pairwise likelihood}

\date{} 					

\author{{Thomas Lumley} \\
	Department of Statistics\\
	University of Auckland\\
	Auckland, New Zealand \\
	\texttt{t.lumley@auckland.ac.nz} \\
	\And
	{Xudong Huang} \\
	Department of Statistics\\
	University of Auckland\\
	Auckland, New Zealand \\
}

\renewcommand{\shorttitle}{\textit{arXiv} Template}

\hypersetup{
pdftitle={Linear mixed models for complex survey data: implementing and evaluating pairwise likelihood},
pdfsubject={stat.ME, stat.CO},
pdfauthor={Thomas Lumley, Xudong Huang},
pdfkeywords={composite likelihood, design-based inference, statistical software},
}

\begin{document}
\maketitle

\begin{abstract}
As complex-survey data becomes more widely used in health and social-science research, there is increasing interest in fitting a wider range of regression models. We describe an implementation of two-level linear mixed models in R using the pairwise composite likelihood approach of Rao and co-workers. We discuss the computational efficiency of pairwise composite likelihood and compare the estimator to the existing stagewise pseudolikelihood estimator in simulations and in data from the PISA educational survey.
\end{abstract}

\keywords{composite likelihood\and design-based inference\and statistical software}

\section{Introduction}
Mixed or multilevel models for variability in regression associations are important in the social sciences and health sciences, so when data are collected using complex survey designs it is of interest to be able to fit these models and estimate the same population parameters as if data were collected from a cohort or with some ignorable sampling design.  Design-based estimation in mixed models is challenging. The classical approach to design-based inference is to use sampling probabilities to reweight estimating functions or pseudolikelihoods that are sums of functions of one observation at a time \citep{binder-ee}.  Since the purpose of mixed models is to estimate relationships between individuals, they intrinsically cannot be reduced to linear estimating functions in this way. 

Two main approaches to design-based inference for mixed models have been proposed.  The first, proposed by \citet{pfeffermann-mixed} and expanded by \citet{rabe-hesketh-survey}  takes advantage of the stagewise independence in stratified multistage sampling.  At each stage, a random sample of next-stage units is taken independently within each stratum, and random effects for each unit are introduced, giving a loglikelihood that is a simple sum and can be reweighted using stage-specific sampling probabilities.    Implementations of this approach and extensions to generalised linear models and other latent variable models, have been developed and are available in standard software \citep{gllamm-framework, mplus, mlwin, stata15}. There does not appear to be a consistent name for this estimation approach; we propose `stagewise pseudolikelihood', emphasising how it takes advantage of stagewise (conditional) independence between clusters at each stage in the design and each level in the model.

The second approach is to replace the population objective function with one that can be more easily estimated.  \citet{rao-mixed} and \citet{yi-rao-mixed} proposed pairwise composite likelihood, where the population objective function is a sum of loglikelihooods for pairs of observations, reweighted using reciprocals of pairwise sampling probabilities. \citet{savitsky-pairwise} also develop a Bayesian approach to inference in a wider range of models based on pairwise likelihood. Software has not previously been available for this approach and, to our knowledge, no comparisons with stagewise pseudolikelihood have been published.   

Each approach has theoretical advantages. Extracting estimates of the realised random effects is straightforward for stagewise pseudolikelihood; these are of interest in themselves and are an important component of efficient quadrature algorithms for generalised linear mixed models.   On the other hand, the stagewise pseudolikelihood approach is applicable only when the clusters in the model are the same as (or nested in) the sampling units in the design, and while the approach allows for stratified sampling, the available Stata implementations do not.  Since the weights do not enter the loglikelihood linearly,  the asymptotic structure for proving consistency of stagewise pseudolikelihood has cluster size as well as cluster number going to infinity, and choices need to be made about the scaling of weights at different stages/levels.

Previous research had applied the pairwise estimator only to the settings where the design and model structure are the same (or nested), but 
in fact the pairwise estimator can be applied to very general designs and models \citep{huang-thesis}.   On the other hand, estimators (predictors) of the random effects are currently not known, and as a result it is not currently possible to use adaptive Gaussian quadrature \citep{gllamm-agq,pinheiro-chao} to fit generalised linear mixed models with pairwise likelihood.  We would also expect some efficiency loss from considering only pairs.

 In this paper we present a novel R \citep{r-itself} implementation of the weighted pairwise likelihood approach, and compare it in simulations to stagewise pseudolikelihood as implemented in Stata \citep{stata15}, and to naive maximum likelihood (which would be appropriate in practice only under ignorable sampling). We consider the setting of a two-level model where the clusters are the same in the design and the model.  The R package is available from \url{https://github.com/tslumley/svylme/tree/pairwise-vs-sequential} and in the supplemental material.

\section{Design and model}
We  consider a two-level linear mixed model as described by \citet{laird-ware-mixed} for groups indexed by $i$ and observations within groups indexed by $j$
$$ Y_{ij} = X_{ij}\beta+Z_{ij}b_i+\epsilon_{ij}$$
where $\epsilon_{ij}\sim N(0,\sigma^2)$, $b_i\sim N(0,\sigma^2V(\theta))$.  In this paper we are interested in estimating $\beta$, $\sigma^2$, and $\theta$, not the realised $b_i$.  

Under this model, $Y|X$ has a multivariate Normal distribution with mean vector $X\beta$. The covariance matrix of $Y$ is block-diagonal with the block for group $i$ being 
$$\sigma^2\Xi_i = \sigma^2(I+Z_i^TVZ_i).$$ 
The loglikelihood for this model is 
$$\ell(\beta,\theta,\sigma^2)= -\frac{1}{2}\sum_i\log |\sigma^2\Xi_i(\theta)| -\frac{1}{2\sigma^2} \sum_i (Y_i-X_i^T\beta)^T\Xi_i(\theta)^{-1}(Y_i-X_i^T\beta)$$

There is a notation clash between the survey literature and the multilevel model literature. Suppose we are studying a sample of students within each of a sample of schools.  From the viewpoint of sampling we would call the schools ``stage 1'' and the students ``stage 2'', but from the viewpoint of multilevel models the students are ``level 1'' and the schools ``level 2''.  We will specify 'stage' or 'level' explicitly.

As noted above, the usual approach to design-based inference is to estimate the population objective function or estimating function by a weighted sum over the sample. It is not straightforward to estimate $\ell$ from a multistage sample: when there is subsampling within groups $\Xi_i^{-1}$ and $|\Xi_i|$ for a group depend on $Z$ for both sampled and non-sampled units.

The population objective function (the `census composite loglikelihood') for the pairwise likelihood approach  would be
$$\tilde \ell_P(\beta,\theta)= \sum_i \sum_{j<k} \ell_{i,jk}(\beta,\theta)$$
where $\ell_{i,jk}(\beta,\theta)$ is the likelihood based on the two observations $Y_{ij}$ and $Y_{ik}$. 
As \citet{lindsay-composite} pointed out when originally describing composite likelihood, each $\ell_{i,jk}(\beta,\theta)$ is a true loglikelihood, so that $E_{\beta_0,\theta_0}[\ell_{i,jk}(\beta,\theta,\sigma^2)]$ is maximised at $(\beta,\theta,\sigma^2)=(\beta_0,\theta_0,\sigma^2_0)$ and the derivative $\nabla\ell_{i,jk}(\beta,\theta,\sigma^2)$ is an unbiased estimating function.  It follows immediately that  $E_{\beta_0,\theta_0,\sigma^2_0}[\tilde \ell_P(\beta,\theta,\sigma^2)]$ is also maximised by $(\beta,\theta,\sigma^2)=(\beta_0,\theta_0,\sigma^2_0)$, and standard smoothness arguments then show that the population maximum pairwise likelihood estimators $(\tilde\beta,\tilde\theta,\tilde\sigma^2)$ are consistent for $(\beta, \theta, \sigma^2)$ as $N\to\infty$.

In a sample we observe only those pairs with $R_{ij}R_{ik}=0$, and need to weight by the reciprocal of the probability of  observing a pair, $\pi_{i,jk}=E[R_{ij}R_{ik}]$, to obtain
$$\hat\ell_P(\beta,\theta,\sigma^2)= \sum_i \sum_{j<k}\frac{R_{i,j}R_{i,k}}{\pi_{i,jk}} \ell_{i,jk}(\beta,\theta,\sigma^2),$$
the design-weighted pairwise loglikelihood.  If the design allows a law of large numbers and central limit theorem, standard arguments again show that the maximum weighted pairwise loglikelihood estimators $(\hat\beta, \hat\theta,\hat\sigma^2)$ are consistent and asymptotically Normal.  Importantly, the asymptotic setting for consistency does not require group sizes to go to infinity.

\section{Computational issues}
\citet{yi-rao-mixed} proposed estimating the parameters by solving the weighted pairwise score equations.  We have done this for simple models, but it is relatively inconvenient to automate for more complex models.  Instead, we follow the approach of \citet{lme4-paper} by profiling out $\beta$ and $\sigma^2$ to obtain a profile weighted pairwise deviance and then using a general-purpose optimiser to minimise it.   Since the dimension of $\theta$ is often much lower than that of $\beta$, profiling gives a simpler optimisation problem.

We define the profile deviance as 
$$\tilde d(\theta) = -2\max_{\beta,\sigma^2} \tilde\ell_P(\beta,\theta,\sigma^2)$$
for the population and 
$$\hat d(\theta) = -2\max_{\beta,\sigma^2} \hat\ell_P(\beta,\theta,\sigma^2)$$
for the sample. 

The corresponding estimates $\tilde\beta_\theta$ and $\hat\beta_\theta$ for $\beta$ are given by generalised least squares in an expanded data set. Define $X_P$ as the $2N_P\times p$ matrix formed by stacking the $2\times p$ design matrices for the $N_p$ pairs.  Similarly, $Y_P$ is formed by stacking the 2-vectors of $Y$ for each pair, and $\Xi_P$ is block-diagonal with $2\times 2$ blocks $\Xi_{i,jk}$.  In the sample, define $X_S$ and $Y_S$ as the rows of $X_P$ and $Y_P$ corresponding to sampled pairs, and $\Xi_S$ as the submatrix of $\Xi_P$ corresponding to sampled pairs.   Note that because $\Xi_P$ is block-diagonal with blocks for each pair, the submatrix of $\Xi_P^{-1}$  corresponding to sampled pairs is just $\Xi_S^{-1}$.

The maximum pairwise likelihood estimate of $\beta$ does not depend on $\sigma^2$, so we can write it just as a function of $\theta$: in the population
$$\tilde\beta_\theta = (X_P^T\Xi_P^{-1}(\theta)X_P)^{-1}X_P^T\Xi_P^{-1}(\theta)Y_P$$
and in the sample
$$\hat\beta_\theta = (X_S^TW^{1/2}\Xi_S^{-1}W^{1/2}(\theta)X_S)^{-1}X_S^TW^{1/2}\Xi_S^{-1}(\theta)W^{1/2}Y_P$$
where $W$ is the diagonal matrix  whose two entries for the pair $(ij, ik)$ are both $\pi_{i,jk}^{-1}$

The MPLE of $\sigma^2$ is 
$$\tilde\sigma^2_\theta =  \frac{1}{2N_P}(Y_P-X_P\tilde\beta_\theta)^T\Xi_P^{-1}(\theta) (Y_P-X_P\tilde\beta_\theta)^T$$
in the population and the weighted estimator in the sample is 
$$\hat\sigma^2_\theta =  \frac{1}{2\hat N_P}(Y_S-X_S\hat\beta_\theta)^T\Xi_S^{-1}(\theta) (Y_S-X_S\hat\beta_\theta)^T$$
where $\hat N_p = \sum_i \sum_{j<k} \pi_{i,jk}^{-1}$.

Plugging these into the  pairwise loglikelihoods gives the population profile pairwise deviance
$$\tilde d(\theta) =  2N_P\log\left(2\pi\tilde\sigma^2_\theta\right) + \sum_{i}\sum_{j<k}\log|\Xi_{i,jk}(\theta)|. $$
and its sample estimator
$$\hat d(\theta) = 2\hat N_P\log\left(2\pi\hat\sigma^2_\theta\right) + \sum_{i}\sum_{j<k}\frac{R_{i,j}R_{i,k}}{\pi_{i,jk}} \log|\Xi_{i,jk}(\theta)|.$$
We use Powell's quadratic bound-constrained optimiser {\tt  BOBYQA} \citep{bobyqa-report} to find $\theta$ minimising $\hat d(\theta)$, with a starting value obtained from an unweighted (`naive') maximum-likelihood fit.  In computing $\hat d(\theta)$ we take advantage of the explicit formulas available for inverse and determinant of $2\times 2$ matrices to rewrite computations over $i,j,k$ and $i',j',k'$ as sets of three or four matrix operations over $i$ and $i'$.

Although there are potentially more pairs than observations, using  pairwise composite likelihood does not significantly increase computational effort, and may actually reduce it when group sizes are large. Consider the unweighted case: if there are $n_1$ groups of size $m$, computing the pairwise profile deviance involves a bounded number  of operations for each of the $m(m-1)$  pairs and so scales as $m^2$. In general, computing the full deviance or score requires computing a determinant and solving an $m\times m$ linear system in each group, both of which scale as $m^3$.  In special cases the determinant may be available in closed form and the matrix $\Xi^{-1}$ for a group may be available directly; computing the deviance then requires an $m\times m$ matrix--vector multiplication, which scales computationally as $m^2$ and is of the same order as the pairwise computation.

\subsection{Variance estimation}
For general designs, a  Horvitz--Thompson-type variance estimator for pairwise composite likelihood estimators involves fourth-order sampling probabilities. These are typically not supplied with survey data. Computing them is straightforward but tedious for simple designs, but is infeasible for the  multi-phase designs used in many large surveys.  It is common practice in secondary analysis of survey data to approximate the variance by treating PSUs as sampled with replacement. Following \citet{yi-rao-mixed} we use a similar with-replacement approximation to make variance calculations tractable.

Define the score for a pair of observations
$$U_{i,jk}=U_{i,jk}(\hat\beta,\hat\theta,\hat\sigma^2)=\left.\frac{\partial \ell_{i,jk}}{\partial\beta}\right|_{(\beta,\theta,\sigma^2)=(\hat\beta,\hat\theta,\hat\sigma^2)}$$
and the corresponding Fisher infomation
$$I_{i,jk}=-\left.\frac{\partial^2 \ell_{i,jk}}{\partial\beta^2}\right|_{(\beta,\theta,\sigma^2)=(\hat\beta,\hat\theta,\hat\sigma^2)}.$$

The empirical population sensitivity and variability matrices \citep{varin-review} are, respectively,
\begin{eqnarray*}
\tilde H & =& \sum_i\sum_{j<k} I_{i,jk}\\
\tilde J &=& \sum_i \sum_{j<k}\sum_{j'<k'}U_{i,jk}^TU_{i,j'k'}\\
\end{eqnarray*}

The variance of the census parameter would be estimated by 
$$\widehat{\mathrm{var}}[\tilde\beta]= \tilde H^{-1}\tilde J\tilde H^{-1}$$

Writing $v^{\otimes 2}$ for $v^Tv$, we approximate the sample sensitivity and variability matrices by
\begin{eqnarray*}
\hat H & =& \sum_i\sum_{j<k} \frac{R_{i,j}R_{i,k}}{\pi_{ij}\pi_{ik}}I_{i,jk}\\
\hat J &=& \frac{n_1}{n_1-1}\sum_i  \left(\frac{R_i}{\pi_i}\sum_{j<k}\frac{R_{i,j}R_{i,k}}{\pi_{j|i}\pi_{k|i}}U_{i,jk}\right)^{\otimes 2}
\end{eqnarray*}
and the variance of $\hat\beta$ by 
$$\widehat{\mathrm{var}}[\hat\beta]= \hat H^{-1}\hat J\hat H^{-1}.$$
Since $\hat\beta$ is a generalised least squares estimator, the form of $\hat H$ and $\hat J$ involve straightforward weighted sums of squares and products of $X$ and residuals.

The same argument can be used to approximate the variances of $\hat\sigma^2$ and $\hat\theta$, but the expressions for $U_{i,jk}$ and $I_{i,jk}$ become more complex, and the Normal approximation to their distribution less accurate. We suggest resampling to estimate the uncertainty in the variance parameters when it is of interest; this is implemented in the \texttt{boot2lme} function.

\subsection{Strata and additional sampling stages}
It is straightforward to allow for additional sampling stages before the stage at which the model groups are sampled. Suppose that a survey takes a stratified sample of school districts, then samples schools within the districts and students within the schools. We can fit a two-level model with schools and students as the levels.  The sampling probabilities $\pi_i$ will be probabilities for schools, and $\pi_{jk|i}$ will be conditional probabilities for pairs of students given schools. Point estimation proceeds exactly as before. 

The only change in variance estimation is that $\hat J$ is replaced by
$$\hat J = \sum_{h\in \mathrm{strata}}\; \frac{n_h}{n_h-1}\sum_{l=1}^{n_h} \left(\sum_{i\in\mathrm{PSU}(l)} \frac{R_i}{\pi_i} \sum_{j<k}\frac{R_{i,j}R_{i,k}}{\pi_{j|i}\pi_{k|i}}U_{i,jk}\right)^{\otimes 2},$$
where $n_h$ is the number of PSUs --- school districts --- in stratum $h$.  That is, the variance is computed treating weighted totals for PSUs in each first-stage stratum as independent and identically distributed, rather than treating weighted totals for groups as independent and identically distributed.

\subsection{User interface}

The {\tt svy2lme} function combines user interface ideas from the {\sf survey} \citep{survey-pkg} and {\sf lme4} \citep{lme4-paper}.  The notation for specifying mixed models is the same as in {\sf lme4}, using the {\tt |} conditioning operator as an addition to the traditional model formula notation.  The survey design is specified using design objects from the {\sf survey} package, which combine the data and survey metadata into a single object; this object is then passed to analysis functions in place of a data frame.

The notation \verb"y~(sex|school)+age+sex" specifies a model with \verb"sex" and \verb"age" and an intercept in the $X$ matrix of fixed effects and \verb"sex" and an intercept in the matrix of random effects $Z$ for each value of \verb"school".   As in {\sf lme4}, the default is that random effects are not constrained to be independent, but this constraint can be added by specifying multiple random-effect groups. That is, \verb"y~(1|school)+(0+sex|school)+age+sex" specifies a random intercept for school and, independent of this, a random coefficient for \texttt{sex} without a random intercept. 

For the common case of two-stage sampling with stratified simple random samples at each stage, the pairwise sampling probabilities can easily be computed from sampling probabilities or population sizes at each stage.  When the user supplies probabilities for each stage but these do not agree with stratified simple random sampling, an approximation due to Haj\'ek (1964) is used within each stage
$$\pi_{ij}\approx \pi_i\pi_j\left[1-\frac{(1-\pi_i)(1-\pi_j)}{\sum_k\pi_k(1-\pi_k)}\right]$$
with the denominator approximated by the unbiased estimator
$$\sum_k  \frac{R_i}{\pi_k}\pi_k(1-\pi_k)=\sum_k  R_i(1-\pi_k).$$

\section{Simulations}
We present simulations comparing the point estimates for $\beta$, $\sigma^2$ and $\theta$ for naive maximum likelihood in the sample, our proposed composite likelihood estimator, and three versions of stagewise pseudolikelihood as implemented in Stata \citep{stata15}. The three stagewise pseudolikelihood estimators are based on three approaches to rescaling the sampling weights to reduce bias: unscaled weights, stage-2 weights scaled to sum to the (population) cluster size, and the proposal of \citet{gk1996} that scales stage-1 weights to the average weight for observations in the cluster and the stage-2 weights to 1.

In the first set of simulations our focus is on the relative efficiency of the composite likelihood estimator.  In the second set, we use simulation settings where the stagewise pseudolikelihood estimator is known to be substantially biased and demonstrate that the composite likelihood estimator does not share its bias.

In each iteration of each simulation we create a finite population satisfying a linear mixed model and  define clusters to be the groups in the mixed model.  We define strata without reference to the model except that clusters are nested in strata, and take a stratified two-stage sample, and then fit the five estimators.  We have one covariate, $X$ with a fixed slope and one covariate, $Z$, with a random intercept and slope. Except as otherwise specified we do not assume the  random intercept and slope are independent (though we do not display the covariance estimate for reasons of space).   That is we simulate from
$$Y=\beta_0+\beta_1 X+\beta_2 X + b_{i0}+b_{iz}Z+\epsilon$$
and fit the model \verb=Y~X+Z+(1+Z | id)=.
All of the simulation code is available at \url{https://github.com/tslumley/svylme-paper} and in the supplementary material. 

 In these simulations we use as the true finite-population parameter values the estimates from a linear mixed model fitted to the whole population and estimate the bias and standard error after subtracting these true values from the estimates. We estimate the bias by the median and the standard error by the scaled median absolute deviation $\mathrm{mad}(x) = 1.4826\times\mathrm{med}|x-\mathrm{med}(x)|.$
 Since the stagewise pseudolikelihood estimator and the pairwise likelihood estimator use essentially the same sandwich variance estimators, and given space restrictions, we do not report a comparison of the estimated standard errors. 

\subsection{Comparing efficiency under non-informative sampling}

In Tables~1--6 we are primarily interested in the efficiency of estimation, comparing stagewise pseudolikelihood, pairwise likelihood, and unweighted (naive) maximum likelihood.  The sampling in all these simulations is non-informative, allowing efficiency to be compared to naive maximum likelihood. 

We see across all the tables that, even under non-informative sampling, stagewise pseudolikelihood without weight scaling is unreliable, giving very large biases for the variance components.  All the other estimators are essentially unbiased for all the parameters in all the settings considered.

stagewise pseudolikelihood with the GK scaling performs very well across these simulations, with essentially no loss of efficiency compared to maximum likelihood.  The cluster size scaling has minor loss of efficiency, primarily  for the variance components. The pairwise likelihood estimator has substantial efficiency loss for the variance components, especially for the random-intercept variance. The loss of efficiency for variance components is larger when the variance components themselves are larger: eg, comparing tables 2 and 5 to 3 and 4. 

Notably, there is much less loss of efficiency for the pairwise likelihood estimator in Table 6, where cluster sizes are smaller.  If all cluster sizes were equal to two and there was simple random sampling of clusters, the pairwise likelihood and maximum likelihood estimators would be the same, so it is reasonable that pairwise likelihood performs relatively better with smaller clusters.

\begin{table}[htp]
\caption{Comparisons from 10,000 replicates. $X$ is an equal mixture of $N(0,1^2)$, $N(0,2^2)$ and $N(0,3^2)$, $Z\sim \Gamma(2)$; cluster size 10; stratum size 1000; sampling 10~clusters per stratum, 2 or 6 observations per cluster}
\centering
\label{table-x-mixture}
\begin{tabular}{lrrrrrr}
\toprule
  &\multicolumn{3}{c}{$\beta$} & \multicolumn{2}{c}{$\mathrm{var}[b]$} & $\sigma^2$\\
 & (Intercept) & z & x & (Intercept) & z & \\ 
\midrule
\multicolumn{6}{l}{\bf Naive ML}  \\~
Bias $\times 100$ & $-$0.2  & 0.0  & 0.0  & $-$1.0  & $-$0.1  & 0.0  \\ 
SE $\times 100$ & 4.3  & 2.2  & 1.0  & 9.8  & 4.9  & 5.1  \\ 
\multicolumn{6}{l}{\bf Pairwise likelihood}  \\
Bias $\times 100$ & $-$0.3  & 0.1  & 0.0  & $-$0.2  & 0.0  & 0.3  \\ 
  SE $\times 100$ & 5.3  & 2.9  & 1.3  & 23.1  & 7.4  & 9.4  \\ 
\multicolumn{6}{l}{\bf SeqPL: cluster size scaling}  \\
Bias $\times 100$ & $-$0.3  & 0.1  & $-$0.1  & $-$0.1  & $-$0.3  & 0.1  \\ 
  SE $\times 100$& 4.9  & 3.1  & 1.1  & 11.4  & 6.1  & 6.4  \\ 
\multicolumn{6}{l}{\bf SeqPL: GK scaling}  \\
Bias $\times 100$ & $-$0.2  & 0.0  & 0.0  & $-$1.3  & $-$0.3  & 0.1  \\ 
  SE $\times 100$ & 4.3  & 2.2  & 1.0  & 9.8  & 4.9  & 5.1  \\ 
\multicolumn{6}{l}{\bf SeqPL: unscaled}  \\
Bias $\times 100$ & $-$0.9  & 0.5  & 0.0  & 851.7  & 303.0  & $-$135.4  \\ 
 SE $\times 100$ & 9.4  & 5.8  & 1.3  & 130.4  & 51.9  & 3.3  \\ 
\bottomrule
\end{tabular}
\end{table}

\begin{table}[htp]
\caption{Comparisons from 10,000 replicates. $X\sim N(0,1)$, $Z\sim \Gamma(2)$;  cluster size 10; stratum size 1000; sampling 10~clusters per stratum, 2 or 6 observations per cluster. DIffers from Table 1 in $X$ distribution}
\centering
\label{table-x-independent}
\begin{tabular}{lrrrrrr}
  \toprule
  &\multicolumn{3}{c}{$\beta$} & \multicolumn{2}{c}{$\mathrm{var}[b]$} & $\sigma^2$\\
 & (Intercept) & z & x & (Intercept) & z & \\ 
  \midrule
\multicolumn{6}{l}{\bf Naive ML}  \\
Bias $\times 100$  & $-$0.2  & 0.1  & $-$0.1  & 0.1  & 0.0  & $-$0.1  \\ 
 SE $\times 100$ & 4.2  & 2.3  & 2.2  & 10.5  & 4.6  & 5.3  \\ 
\multicolumn{6}{l}{\bf Pairwise likelihood}  \\
  Bias $\times 100$  & $-$0.1  & 0.0  & $-$0.2  & $-$1.6  & $-$0.7  & 0.1  \\ 
  SE $\times 100$ & 5.7  & 3.2  & 3.0  & 23.0  & 7.2  & 9.4  \\ 
 \multicolumn{6}{l}{\bf SeqPL: cluster size scaling}  \\
 Bias $\times 100$  & 0.0  & 0.0  & $-$0.2  & $-$0.4  & $-$0.3  & $-$0.3  \\ 
  SE $\times 100$ & 5.4  & 3.3  & 2.7  & 11.4  & 6.2  & 6.3  \\ 
 \multicolumn{6}{l}{\bf SeqPL: GK scaling}  \\
 Bias $\times 100$  & $-$0.2  & 0.1  & $-$0.1  & $-$0.2  & $-$0.1  & 0.0  \\ 
  SE $\times 100$ & 4.2  & 2.3  & 2.2  & 10.5  & 4.6  & 5.3  \\ 
\multicolumn{6}{l}{\bf SeqPL: unscaled}  \\
  Bias $\times 100$  & $-$0.3  & 0.2  & $-$0.1  & 857.5  & 305.6  & $-$135.3  \\ 
SE $\times 100$ & 9.2  & 5.6  & 2.8  & 129.5  & 53.2  & 3.4  \\ 
   \bottomrule
\end{tabular}
\end{table}

\begin{table}[htp]
\caption{Comparisons from 10,000 replicates. $X\sim N(0,1)$, $Z\sim \Gamma(2)$;  cluster size 10; stratum size 1000; sampling 10~clusters per stratum, 2 or 6 observations per cluster. This differs from Table~2  in that random effect variances are smaller. }
\centering
\label{table-weaker}
\begin{tabular}{lrrrrrr}
  \toprule
  &\multicolumn{3}{c}{$\beta$} & \multicolumn{2}{c}{$\mathrm{var}[b]$} & $\sigma^2$\\
 & (Intercept) & z & x & (Intercept) & z & \\ 
  \midrule
\multicolumn{6}{l}{\bf Naive ML}  \\
Bias $\times 100$  & 0.2  & 0.0  & 0.0  & $-$0.6  & 0.0  & 0.1  \\ 
 SE $\times 100$ & 3.5  & 1.5  & 1.9  & 7.3  & 1.1  & 5.2  \\ 
 \multicolumn{6}{l}{\bf Pairwise likelihood}  \\
 Bias $\times 100$  & 0.3  & $-$0.1  & $-$0.1  & $-$0.1  & 0.0  & $-$0.2  \\ 
 SE $\times 100$ & 4.2  & 1.9  & 2.4  & 11.7  & 1.6  & 7.2  \\ 
 \multicolumn{6}{l}{\bf SeqPL: cluster size scaling}  \\
 Bias $\times 100$  & 0.4  & $-$0.1  & $-$0.2  & $-$0.3  & $-$0.1  & $-$0.3  \\ 
  SE $\times 100$ & 4.2  & 1.9  & 2.4  & 8.2  & 1.3  & 6.2  \\ 
  \multicolumn{6}{l}{\bf SeqPL: GK scaling}  \\
Bias $\times 100$  & 0.2  & 0.0  & $-$0.1  & $-$0.7  & $-$0.1  & 0.1  \\ 
  SE $\times 100$ & 3.5  & 1.6  & 1.9  & 7.3  & 1.1  & 5.2  \\ 
 \multicolumn{6}{l}{\bf SeqPL: unscaled}  \\
Bias $\times 100$  & 0.4  & $-$0.2  & 0.0  & 670.4  & 204.2  & $-$133.1  \\ 
  SE $\times 100$ & 7.9  & 4.5  & 2.8  & 95.4  & 34.0  & 3.9  \\ 
   \bottomrule
\end{tabular}
\end{table}

\begin{table}[htp]
\caption{Comparisons from 10,000 replicates. $X$ is correlated within clusters--- the sum of independent $N(0,1)$ and cluster-specific $N(0,1)$. $Z\sim \Gamma(2)$;  cluster size 10; stratum size 1000; sampling 10~clusters per stratum, 2 or 6 observations per cluster. Differs from Table~3 only in the correlation for $X$ }
\centering

\label{table-correlated-x}
\begin{tabular}{lrrrrrr}
  \toprule
  &\multicolumn{3}{c}{$\beta$} & \multicolumn{2}{c}{$\mathrm{var}[b]$} & $\sigma^2$\\
 & (Intercept) & z & x & (Intercept) & z & \\ 
  \midrule
\multicolumn{6}{l}{\bf Naive ML}  \\
Bias $\times 100$ & 0.1  & 0.0  & 0.1  & 0.2  & 0.0  & $-$0.4  \\ 
   SE $\times 100$ & 3.3  & 1.4  & 1.4  & 6.7  & 1.1  & 5.1  \\ 
\multicolumn{6}{l}{\bf Pairwise likelihood}  \\
Bias $\times 100$ & $-$0.1  & 0.0  & 0.0  & 0.2  & 0.0  & $-$0.3  \\ 
   SE $\times 100$ & 4.6  & 1.8  & 1.8  & 11.5  & 1.7  & 7.2  \\ 
 \multicolumn{6}{l}{\bf SeqPL: cluster size scaling}  \\
 Bias $\times 100$ & $-$0.1  & 0.0  & 0.0  & 0.3  & $-$0.1  & $-$0.2  \\ 
   SE $\times 100$ & 4.6  & 1.8  & 1.8  & 7.9  & 1.3  & 5.9  \\ 
   \multicolumn{6}{l}{\bf SeqPL: GK scaling}  \\
Bias $\times 100$ & 0.2  & 0.0  & 0.1  & 0.0  & 0.0  & $-$0.3  \\ 
   SE $\times 100$ & 3.3  & 1.4  & 1.4  & 6.7  & 1.1  & 5.1  \\ 
 \multicolumn{6}{l}{\bf SeqPL: unscaled}  \\
Bias $\times 100$ & 0.1  & 0.0  & 0.1  & 671.7  & 200.7  & $-$132.9  \\ 
   SE $\times 100$ & 8.6  & 4.9  & 2.6  & 88.0  & 32.0  & 3.5  \\ 
   \bottomrule
\end{tabular}
\end{table}

\begin{table}[htp]
\caption{Comparisons from 10,000 replicates. $X$ is correlated within clusters--- the sum of independent $N(0,1)$ and cluster-specific $N(0,1)$. $Z\sim \Gamma(2)$;  cluster size 10; stratum size 1000; sampling 10~clusters per stratum, 2 or 6 observations per cluster. Differs from Table~4 in having larger random-effect variances ($\sigma=1$)}
\centering
\label{tabel-bigger-effects}
\begin{tabular}{lrrrrrr}
  \toprule
  &\multicolumn{3}{c}{$\beta$} & \multicolumn{2}{c}{$\mathrm{var}[b]$} & $\sigma^2$\\
 & (Intercept) & z & x & (Intercept) & z & \\ 
  \midrule
\multicolumn{6}{l}{\bf Naive ML}  \\
Bias $\times 100$ & 0.2  & $-$0.1  & 0.1  & 0.2  & 0.1  & 0.0  \\ 
  SE $\times 100$ & 4.1  & 2.3  & 1.8  & 12.7  & 4.6  & 5.3  \\ 
 \multicolumn{6}{l}{\bf Pairwise likelihood}  \\
 Bias $\times 100$ & 0.1  & 0.0  & 0.1  & $-$1.5  & $-$0.6  & $-$0.1  \\ 
  SE $\times 100$ & 5.8  & 3.2  & 2.6  & 23.9  & 7.2  & 9.5  \\ 
  \multicolumn{6}{l}{\bf SeqPL: cluster size scaling}  \\
 Bias $\times 100$ & 0.1  & $-$0.1  & 0.0  & $-$0.2  & $-$0.4  & 0.1  \\ 
  SE $\times 100$ & 5.5  & 3.3  & 2.4  & 14.1  & 5.8  & 6.1  \\ 
  \multicolumn{6}{l}{\bf SeqPL: GK scaling}  \\
  Bias $\times 100$ & 0.2  & $-$0.1  & 0.1  & $-$0.2  & 0.0  & 0.1  \\ 
  SE $\times 100$ & 4.1  & 2.3  & 1.8  & 12.7  & 4.6  & 5.3  \\ 
 \multicolumn{6}{l}{\bf SeqPL: unscaled}  \\
 Bias $\times 100$ & $-$0.2  & $-$0.1  & 0.2  & 885.7  & 311.4  & $-$135.5  \\ 
   SE $\times 100$ & 10.0  & 6.3  & 2.7  & 126.0  & 51.6  & 3.3  \\ 
   \bottomrule
\end{tabular}
\end{table}

\begin{table}[htp]
\caption{Comparisons from 10,000 replicates. $X\sim N(0,1)$, $Z\sim \Gamma(2)$;   stratum size 1000; population cluster size 5, sample cluster size 3, sample 20~clusters per stratum. Model assumes independence of the intercept and slope random effects: {\tt Y\textasciitilde X+Z+(1 | id)+(0+Z | id)}}
\centering
\label{table-small-clusters}
\begin{tabular}{lrrrrrr}
  \toprule
  &\multicolumn{3}{c}{$\beta$} & \multicolumn{2}{c}{$\mathrm{var}[b]$} & $\sigma^2$\\
 & (Intercept) & z & x & (Intercept) & z & \\ 
  \midrule
\multicolumn{6}{l}{\bf Naive ML}  \\
Bias $\times 100$ & $-$0.1  & 0.0  & 0.0  & 0.1  & 0.0  & 0.0  \\ 
   SE $\times 100$ & 2.4  & 0.4  & 0.5  & 3.5  & 0.1  & 0.1  \\ 
 \multicolumn{6}{l}{\bf Pairwise likelihood}  \\
 Bias $\times 100$ & 0.0  & 0.0  & 0.0  & 0.0  & 0.0  & 0.0  \\ 
  SE $\times 100$ & 2.4  & 0.4  & 0.5  & 3.5  & 0.1  & 0.2  \\ 
  \multicolumn{6}{l}{\bf SeqPL: cluster size scaling}  \\
 Bias $\times 100$ & $-$0.1  & 0.0  & 0.0  & 0.0  & 0.0  & 0.0  \\ 
  SE $\times 100$ & 2.4  & 0.4  & 0.5  & 3.5  & 0.1  & 0.1  \\ 
   \multicolumn{6}{l}{\bf SeqPL: GK scaling}  \\
 Bias $\times 100$ & $-$0.1  & 0.0  & 0.0  & 0.0  & 0.0  & 0.0  \\ 
   SE $\times 100$ & 2.4  & 0.4  & 0.5  & 3.5  & 0.1  & 0.1  \\ 
 \multicolumn{6}{l}{\bf SeqPL: unscaled}  \\
  Bias $\times 100$ & $-$0.1  & 0.0  & 0.0  & 7.2  & 2.4  & $-$2.3  \\ 
  SE $\times 100$ & 2.4  & 0.6  & 0.5  & 3.7  & 0.3  & 0.1  \\ 
   \bottomrule
\end{tabular}
\end{table}
\subsection{Comparing bias under strongly informative sampling}

The stagewise pseudollikelihood estimator is consistent as cluster size goes to infinity, regardless of the sampling design, and without need for weight scaling, because the realised random effects can then be estimated precisely and the estimation problem effectively reduces to linear regression with cluster-specific offsets.  When clusters are not large, the stagewise pseudolikelihood estimator can be biased, especially when sampllng is strongly informative for the random effects, and the weight scaling strategy matters.  In contrast, the weighted pairwise score equations are unbiased under any sampling design. 

We present three simulations with strongly informative sampling to illustrate this distinction.  They use the same population, strata, and clusters as Table 2, but differ in the sampling.  In Table~\ref{info1}, 2 or 6 observations are taken from a cluster according to whether the residual variance is above the median, in Table~\ref{info2}, 2 or 6 observations are taken according to whether the absolute value of $b_1$ is above the median, and in Table~\ref{info3}, 2 or 6 observations are taken according to whether $b_1$ is positive or negative.

In all three scenarios, the pairwise likelihood estimator is approximately unbiased for all parameters. It still has higher variance than the other estimators, as it did under non-informative sampling.  stagewise pseudolikelihood with unscaled weights is approximately unbiased for the regression parameters, but as before has substantial bias for the variance parameters.  Each of the scaled stagewise pseudolikelihood estimators fails for at least one of the scenarios, and in Table 9 the two scaled estimators have appreciable bias even in the regression coefficients.

\begin{table}[htp]
\caption{Comparisons from 10,000 replicates. $X\sim N(0,1)$, $Z\sim \Gamma(2)$;  cluster size 10; stratum size 1000; sampling 10~clusters per stratum, 2 or 6 observations per cluster. Two observations from clusters where $\mathrm{var}[\epsilon]$ is below its median and six where $\mathrm{var}[\epsilon]$ is above its median.}
\centering
\label{info1}
\begin{tabular}{lrrrrrr}
  \toprule
  &\multicolumn{3}{c}{$\beta$} & \multicolumn{2}{c}{$\mathrm{var}[b]$} & $\sigma^2$\\
 & (Intercept) & z & x & (Intercept) & z & \\ 
  \midrule
\multicolumn{6}{l}{\bf Naive ML}  \\
Bias $\times 100$ & 0.1  & 0.0  & 0.2  & $-$12.7  & $-$3.2  & 54.5  \\ 
  SE $\times 100$ & 4.3  & 2.4  & 2.4  & 10.9  & 4.6  & 5.4  \\ 
 \multicolumn{6}{l}{\bf Pairwise likelihood}  \\
 Bias $\times 100$ & 0.1  & $-$0.1  & 0.2  & $-$0.7  & $-$0.5  & 0.0  \\ 
  SE $\times 100$ & 5.0  & 3.0  & 2.9  & 18.7  & 5.7  & 6.6  \\ 
  \multicolumn{6}{l}{\bf SeqPL: cluster size scaling}  \\
 Bias $\times 100$ & 0.0  & $-$0.1  & 0.2  & $-$15.0  & $-$4.8  & 21.6  \\ 
  SE $\times 100$ & 4.6  & 3.0  & 2.6  & 8.9  & 5.3  & 5.1  \\ 
   \multicolumn{6}{l}{\bf SeqPL: GK scaling}  \\
 Bias $\times 100$ & 0.1  & 0.0  & 0.2  & $-$13.0  & $-$3.4  & 54.6  \\ 
  SE $\times 100$ & 4.3  & 2.4  & 2.4  & 10.8  & 4.6  & 5.4  \\ 
 \multicolumn{6}{l}{\bf SeqPL: unscaled}  \\
  Bias $\times 100$ & 0.2  & 0.0  & $-$0.1  & 423.5  & 136.3  & $-$105.8  \\ 
  SE $\times 100$ & 6.5  & 3.9  & 2.9  & 42.6  & 16.8  & 3.1  \\ 
   \bottomrule
\end{tabular}
\end{table}

\begin{table}[htp]
\caption{Comparisons from 10,000 replicates. $X\sim N(0,1)$, $Z\sim \Gamma(2)$;  cluster size 10; stratum size 1000; sampling 10~clusters per stratum, 2 or 6 observations per cluster. Two observations from clusters where $|b_1|$ is below the median and six  where $|b_1|$ is above the median.}
\centering
\label{info2}
\begin{tabular}{lrrrrrr}
  \toprule
  &\multicolumn{3}{c}{$\beta$} & \multicolumn{2}{c}{$\mathrm{var}[b]$} & $\sigma^2$\\
 & (Intercept) & z & x & (Intercept) & z & \\ 
  \midrule
\multicolumn{6}{l}{\bf Naive ML}  \\
Bias $\times 100$ & 0.0  & 0.0  & 0.0  & $-$1.2  & 20.8  & $-$2.9  \\ 
   SE $\times 100$ & 4.1  & 2.2  & 2.2  & 10.3  & 3.6  & 5.0  \\ 
\multicolumn{6}{l}{\bf Pairwise likelihood}  \\
 Bias $\times 100$ & $-$0.3  & 0.1  & 0.1  & $-$0.6  & 0.0  & 0.3  \\ 
  SE $\times 100$ & 5.3  & 2.8  & 3.0  & 21.0  & 5.3  & 9.3  \\ 
 \multicolumn{6}{l}{\bf SeqPL: cluster size scaling}  \\
  Bias $\times 100$ & $-$0.1  & 0.1  & 0.1  & $-$1.5  & $-$20.0  & $-$6.0  \\ 
 SE $\times 100$ & 5.0  & 3.0  & 2.6  & 10.6  & 3.3  & 5.8  \\ 
  \multicolumn{6}{l}{\bf SeqPL: GK scaling}  \\
  Bias $\times 100$ & 0.0  & 0.0  & 0.0  & $-$1.5  & 20.6  & $-$2.8  \\ 
  SE $\times 100$ & 4.1  & 2.2  & 2.2  & 10.2  & 3.6  & 5.1  \\ 
 \multicolumn{6}{l}{\bf SeqPL: unscaled}  \\
  Bias $\times 100$ & 0.0  & 0.2  & $-$0.1  & 858.6  & 310.2  & $-$135.7  \\ 
   SE $\times 100$ & 9.7  & 5.5  & 2.7  & 128.8  & 56.1  & 3.3  \\ 
   \bottomrule
\end{tabular}
\end{table}

\begin{table}[htp]
\caption{Comparisons from 10,000 replicates. $X\sim N(0,1)$, $Z\sim \Gamma(2)$;  cluster size 10; stratum size 1000; sampling 10~clusters per stratum, 2 or 6 observations per cluster. Two observations were sampled from clusters where $b_z<0$ and six where $b_z\geq0$}
\centering
\label{info3}
\begin{tabular}{lrrrrrr}
  \toprule
  &\multicolumn{3}{c}{$\beta$} & \multicolumn{2}{c}{$\mathrm{var}[b]$} & $\sigma^2$\\
 & (Intercept) & z & x & (Intercept) & z & \\ 
  \midrule
\multicolumn{6}{l}{\bf  Naive ML}  \\
Bias $\times 100$ & $-$14.8  & 15.7  & $-$0.1  & $-$0.1  & $-$8.5  & 0.6  \\ 
 SE $\times 100$ & 4.3  & 2.4  & 2.2  & 10.0  & 5.4  & 5.6  \\ 
\multicolumn{6}{l}{\bf  Pairwise likelihood}  \\
  Bias $\times 100$ & 0.1  & 0.0  & 0.1  & 0.2  & $-$0.4  & $-$0.3  \\ 
 SE $\times 100$ & 5.3  & 3.0  & 3.3  & 24.1  & 7.0  & 9.8  \\ 
 \multicolumn{6}{l}{\bf  SeqPL: cluster size scaling}  \\
  Bias $\times 100$ & $-$14.0  & $-$26.0  & 0.0  & $-$2.8  & $-$6.9  & $-$2.7  \\ 
 SE $\times 100$ & 6.3  & 3.3  & 2.8  & 10.9  & 5.4  & 6.7  \\ 
  \multicolumn{6}{l}{\bf  SeqPL: GK scaling}  \\
  Bias $\times 100$ & $-$14.8  & 15.7  & $-$0.1  & $-$0.4  & $-$8.7  & 0.7  \\ 
 SE $\times 100$ & 4.3  & 2.4  & 2.2  & 10.0  & 5.4  & 5.6  \\ 
 \multicolumn{6}{l}{\bf  SeqPL: unscaled}  \\
  Bias $\times 100$ & $-$2.1  & 3.8  & 0.0  & 862.4  & 304.0  & $-$135.7  \\ 
 SE $\times 100$ & 9.4  & 5.7  & 2.7  & 138.8  & 56.0  & 4.4  \\ 
   \bottomrule
\end{tabular}
\end{table}

\section{Example}

We fit a linear mixed model to data selected from the PISA 2012 educational survey \citep{pisa2012}, obtained from the {\sf pisa2012lite} R package\citep{pisa2012lite-pkg}. PISA is an international survey including questionnaires about school, parent, and student characteristics and evaluations of student performance on a variety of domains, with subsampling and multiple imputation to reduce respondent burden.   We analysed data on mathematics performance and gender, related problem-solving skills, the proportion of girls at the school, and the ratio of students to mathematics teachers, using the New Zealand subset of the data.  The outcome variable is provided as five multiple imputations (`plausible values'), and we present results for the first plausible value from  weighted pairwise likelihood and from Stata's stagewise pseudolikelihood using the `gk' scaling. We also present a combined result from all five plausible values using weighted pairwise likelihood and combining results with Rubin's rules \citep{rubin-rules}.  The full code and results are in the supplementary material and at \url{github.com/tslumley/svylme-paper}.  The stagewise pseudolikelihood estimator was very close to the boundary of the parameter space, and different maximisation options produced somewhat different variance component estimates. 

\begin{table}
\caption{Analysis of NZ maths performance data and gender, comparing pairwise and stagewise pseudolikelihood: first plausible value only. Standard errors from sandwich estimator for stagewise pseudolikelihood and from both sandwich and bootstrap for pairwise estimator. }
\label{pv1math}
\centering
\begin{tabular}{lrr|rrr}
& \multicolumn{2}{c}{stagewise} & \multicolumn{3}{c}{Pairwise}\\
Parameter & $\hat\beta$ & SE & $\hat\beta$ & sand. & boot.\\
\hline
(Intercept)  & 467 & 19  &      490 &17 &20 \\
Male    & 50   &  21 &  69& 21 &22\\
Prop. girls &60 & 16   &               54 &  16 & 19\\
staff:student ratio& 0.06 & 0.12&  -0.03 &   0.09 & 0.11\\
maths self-efficicacy& 40 & 2.3&               47 &   2.0 & 2.2\\ 
openness   &  17 & 2.2    &         14 &  2.6& 2.9\\
Male: Prop. girls & -97 & 28& -129 &  33 & 37 \\
male: staff-student ratio & 0.00 & 0.11 &  -0.07 & 0.11 & 0.14\\
\hline
residual variance& 4901& 165 & 4830&---&227 \\
intercept & 803 & 484 & 536 &---& 267\\
cov & 152 & 87 & 250&---&135\\
gender& 29 & 45 &119&---&175\\
\hline
\end{tabular}

\end{table}

Table~\ref{pv1math} shows that the two estimators give broadly comparable results. Since `female' is the reference category for the gender variable, the interaction with proportion of girls shows that girls did better in schools with more girls and boys did better in schools with more boys. Mathematics self-efficacy and openness to problem solving are strongly associated with better results. Staff:student ratio shows no evidence of association.  There is modest evidence of variation in the mean result between schools, which could be quite large.  Variation in the gender difference appears to be small, and may be essentially non-existent. 

The sandwich standard errors are 10--20\% smaller than the bootstrap standard errors; the bootstrap is recommended if it is feasible. 

Table~\ref{allmath} reinforces these messages.  As the higher standard deviations for the variance components indicate, the variance component estimates were less stable between plausible values than the fixed-effect estimates.

\begin{table}
\caption{Analysis of NZ maths performance data and gender, with the pairwise likelihood estimator based on all five  plausible values. Standard errors are shown using both sandwich estimator and bootstrap. }
\label{allmath}
\centering

\begin{tabular}{lrrr}
&  \multicolumn{3}{c}{Pairwise}\\
Parameter &  $\hat\beta$ & sandwich & bootstrap\\
\hline
(Intercept)  & 495 & 18 & 21\\
Male    & 63 & 22 & 23\\
Prop. girls & 48 & 17 & 20\\
staff:student ratio& -0.04 & 0.10 & 0.12\\
maths self-efficicacy& 47 & 2.4 & 2.7\\
openness   & 13 & 3.3 & 3.7 \\
Male$\times$ Prop. girls &  -122 & 35 & 38\\
male$\times$ staff-student ratio & -0.05 & 0.12 & 0.14\\
\hline
residual variance& 4940 & --- & 292\\
intercept & 656 & --- & 354\\
cov & 142 & --- & 202 \\
gender& 131 & --- & 223\\
\hline
\end{tabular}

\end{table}

\section{Discussion}

The pairwise likelihood estimator is less efficient than the stagewise pseudolikelihood estimator, and while it is more widely reliable, the settings where informative sampling causes bias in the stagewise pseudolikelihood estimator are arguably unrealistic in any practical application.  Our results confirm again that appropriate scaling of weights at each stage is important for stagewise pseudolikelihood.

It may be surprising that the pairwise likelihood estimator can be so inefficient, since Normal distributions are characterised by their means and variances.  There are two potential explanations.  First, the loglikelihoods for pairs are correlated and we do not take advantage of this correlation.   Second, the Gaussian loglikelihood depends more directly on the precision matrix  than the variance matrix, and the sample precision matrix is not the sample submatrix of the population precision matrix.

The pairwise likelihood estimator has the potential to be extended to settings where the model groups and design clusters are independent.  Its performance compared to the stagewise pseudolikelihood estimator is sufficiently good that such an extension would be worth pursuing.

\bibliographystyle{unsrtnat}
\bibliography{svylme}  






\end{document}